# Scaling Digital Twin Models

Deniz Karanfil [a,b,c,d], Bahram Ravani [a,b,e]

[a] *Department of Mechanical and Aerospace Engineering, University of California, Davis, CA 95616, USA*
[b] *Advanced Highway Maintenance and Construction Technology (AHMCT) Research Center at the University of California, Davis, CA 95616, USA*
[c] *dkaranfil@ucdavis.edu (Corresponding author)*
[d] *This paper presents research conducted as part of the first author's Ph.D. dissertation entitled Developing Scalable Digital Twins of Construction Vehicles.*
[e] *bravani@ucdavis.edu*

## ABSTRACT

In many industries, the scale and complexity of systems can present significant barriers to the development of accurate digital twin models. This paper introduces a novel methodology and a modular computational tool utilizing machine learning and dimensional analysis to establish a framework for scaling digital twin models. Scaling techniques have not yet been applied to digital twin technology, but they can eliminate the need for repetitive physical calibration of such models in industries where product lines include a variety of sizes of the same or similar products. In many cases, it may be easier or more cost-effective to perform physical calibration of the digital twin model on smaller units of a product line. Scaling techniques can then allow adapting the calibration data from the smaller units to other sizes of the product line without the need for additional data collection and experimentation for calibration. Conventional application of dimensional analysis for scaling in this context introduces several challenges due to distortion of scaling factors. This paper addresses these challenges and introduces a framework for proper scaling of digital twin models. The results are applied to scaling the models between an industrial-size wheel loader vehicle used in construction to a miniaturized system instrumented in a laboratory setting.

## 1. Introduction

Digital twins are becoming increasingly integrated across a wide range of industries. Notably, in the construction sector their adoption is gaining momentum [1-4], which is a field that suffers significantly from low productivity in the execution of repetitive tasks [5-7]. In the context of construction vehicles, there has been previous research focusing on the development of physics-based digital twins for systems such as excavators [8,9], tower cranes [10], and wheel loaders [11,12]. However, validating the digital twin models in cases with complex underlying physics such as the ones involving interactions of the construction equipment with the environment requires extensive data collection and calibration of the digital models. Generating calibration data for such systems can be both costly and logistically challenging, primarily due to the difficulty in installing in-situ sensors for proper data collection. An alternative methodology proposed in this paper involves instrumenting, for example, a small-scale system within a similar product line and scaling the results to generate calibration data for variables that are costly or challenging to measure directly in the large-scale counterpart.

To scale a digital twin model, it is essential to utilize dimensionless numbers. The foundational ideas of physical dimensions and dimensional analysis were first introduced by Joseph Fourier and Lord Rayleigh respectively [13-16]. Scaling laws, which are relationships derived from dimensionless numbers, provide a systematic framework for predicting system behavior across different scales. Scaling facilitates the design and analysis of complex and large-scale physical systems by reducing reliance on extensive computational resources or large-scale experimental setups that increase costs and limit feasibility [17,18]. Dimensional analysis is vital for developing scaling models because it reduces the complexity of the problem by reducing the number of parameters involved in it [19]. The Buckingham Pi theorem [20] is a fundamental tool in dimensional analysis, and it states that systems can be modeled with a reduced number of parameters using dimensionless numbers. The theorem also facilitates the systematic derivation of the dimensionless numbers of physical systems, which make the analyses of systems more robust by aiding them when it comes to experimental design, data correlation, and result interpretation. It also enables the derivation of scaling laws for systems where similitude requirements hold true [21,22].

Previous studies in similitude analysis have explored methodologies and algorithms capable of automatically generating dimensionless numbers via the Buckingham Pi theorem [23]. Although the Buckingham Pi theorem can be employed to derive these terms, using them directly to scale physical parameters poses critical limitations. Namely, the application of the theorem does not result in unique solutions (dimensionless numbers) [17]. Furthermore, the theorem on its own is inadequate to discern which dimensionless terms hold higher physical significance [24]. The terms with lower physical significance may offer limited utility of the scaling process if they do not exhibit strong correlations with the parameters intended to scale. Xie et al. [25] have focused on identifying the significant dimensionless numbers from data by employing machine learning and data-driven analysis. Although more significant dimensionless numbers can be identified, a notable gap in the literature remains on the impact of distorted scaling factors on how the parameters of interest scale. Established scaling laws can break down even with minor distortions to the scaling factors as these distortions alter the underlying relationships between the dimensionless numbers. While there have been previous studies where the conventional Pi theorem–based scaling approach was employed for complex mechanical systems (see, for example, Zhao et al. [26]), the application of scaling laws necessitates that no distortion is present in any of the scaling factors between the small-scale and the large-scale systems [27,28]. This is a condition that is unlikely to be met in real-world settings involving the use of existing product lines. When an existing small-scale unit is employed, distortions in scaling factors are almost inevitable, which would invalidate the derived scaling laws. However, if the issue of distortions can be effectively addressed, using a small-scale unit of a product line to generate calibration data for the digital twin of a large-scale product from the same line becomes significantly more feasible. This paper develops a methodology which enables the use of existing units within a product line as sources of training data for a model that can produce accurate estimations for variables that are costly to measure across a range of scales, even in the presence of distorted scaling factors. In handling the distortions, the methodology developed here uses a machine learning-based algorithm that incorporates dimensionless parameters associated with the system, as well as the influence of distorted scaling factors.

Previous literature in modeling the relationship between the dimensionless terms so that the parameter of interest can be scaled includes the work of Kasprzak et al. [29]. Regression methods and a variety of modeling assumptions are often employed to capture relationships among dimensionless terms [30]. For instance, the algorithms proposed by Mendez and Ordonez utilize power law formulations in this context [31,32]. However, such power law formulation only applies to a limited class of problems. In the context of scaling calibrated digital twin models, the effect of distortions must also be included in the model. Neural networks often offer higher versatility compared to traditional regression methods when it comes to complex and highly nonlinear relationships [33]. In this paper, neural networks are employed to estimate critical variables needed for calibration across different scales even when distortions are present in the scaling factors. This is achieved through the involvement of the distortions and their effects on scaling through prediction factors [17]. In similitude analysis, the small-scale system is often referred to as the model and the larger-scale system as the prototype. A prediction factor refers to the ratio of the dimensionless numbers of the model and the prototype which is not equal to one due to scaling distortions [17]. This paper addresses gaps in existing literature by employing machine learning to model the effects of various distortions on the prediction factors, thereby enabling parameter scaling.

This approach facilitates the development of a digital twin for the prototype at lower costs using fewer sensors. The efficacy of the scaling framework developed in this paper is demonstrated by applying it to scale bucket forces in excavation using a wheel loader vehicle determined from a calibrated physics-based digital twin of an industrial-size wheel loader to accurately estimate the forces acting on a miniature wheel loader.

## 2. The Scaling Framework

*2.1. Calibrated Digital Twin of the Wheel Loader*

The digital twin employed in this study is a physics-based digital twin of a commercial wheel loader. Its development involved the integration of a high-fidelity CAD model into physics-based multibody dynamics simulations using the software Algoryx Dynamics [34,35]. The main purpose of this digital twin is to accurately estimate the forces acting on the bucket of the wheel loader during excavation of a pile of soil or rocks which has been used for tasks such as soil parameter estimation [36] and powertrain control [37]. This digital twin was calibrated using real-world experimental data gathered by instrumenting the commercial wheel loader with a wide array of sensors including an inclinometer, a quadrature encoder, pressure transducers, and load pins mounted at the joints between the base of the bucket and the remaining links to measure forces acting on the bucket, as illustrated in Fig. 1 [12]. The idea here is to scale this model to develop digital twins of other models of the wheel loader of different sizes without repeating the calibration process. In this fashion, much of the cost and effort in instrumentation and experimentation for calibration of other sizes of the wheel loader product line can be eliminated.

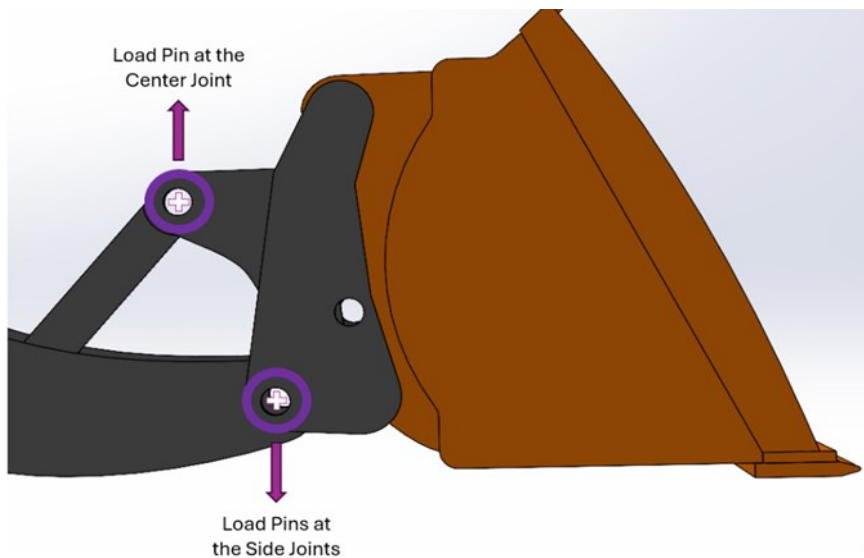

**Fig. 1.** Hinges where the load pins were mounted.

The commercial wheel loader was instrumented with load pins in the side and the center joints of its bucket (see Fig. 1) to obtain experimental estimates of the excavation forces. If estimates of excavation forces from an instrumented wheel loader can be scaled to other sizes of such vehicles without the need to instrument them, then scaling provides much savings in cost and effort. This is the hypothesis which is proved in this paper.

In scaling, it is essential to identify which parameters influence the target parameter being scaled, based on the physics of the system. Section 2.3 explains this process in detail. The bucket lifting mechanism of the commercial wheel loader used is shown in Fig. 2. The components of the bucket lift mechanism are numbered as listed in Table 1.

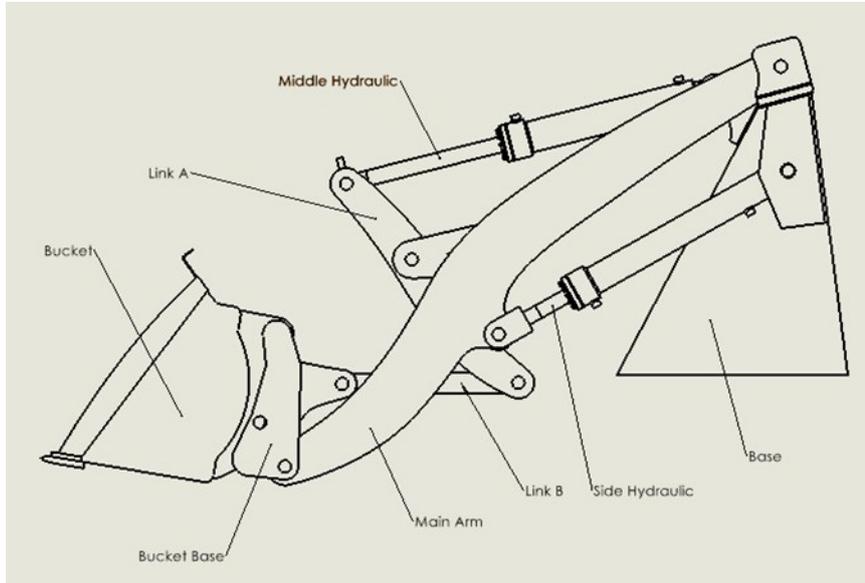

**Fig. 2.** Names of the mechanism components shown on the technical drawing.

**Table 1**
Indexing of the mechanism components.

| Part | Body Index |
| --- | --- |
| Bucket | 1 |
| Link B | 2 |
| Main Arm | 3 |
| Base | 4 |

*2.2. Determination of the Number of Input Parameters in the Scaling Framework*

The Buckingham Pi theorem [20] is used to determine the number of dimensionless numbers required to express the model of the physical system in a dimensionless form. The dimensionless form results in reduced number of parameters which is expressed by Eq. (1) [20].

$$n = p - f \quad (1)$$

where $p$ denotes the number of parameters in the original form of a physical system, $n$ represents the number of parameters in the dimensionless form, and $f$ corresponds to the number of fundamental units involved. Fundamental units, also referred to as fundamental dimensions, are the units which cannot be expressed in terms of other units [17]. The fundamental units of a system are determined by the energy domains (mechanical, electrical, etc.) involved in the physical system under consideration. How many fundamental units a system has ($f$) can be found as shown in Eq. (2) [38].

$$f = \sum_{1}^{i} \beta_i - \sum_{2}^{i} \xi_{1i} \quad (2)$$

where $i$ refers to the number of energy domains present in the system. $\beta_i$ denotes how many fundamental units there are within domain $i$. $\xi_{1i}$ represent the number of fundamental units the domains 1 and $i$ have in common. The bucket lift mechanism of a wheel loader has hydraulic and mechanical systems, which share all three of their fundamental units (kilograms, meters, and seconds representing mass, length, and time, respectively). Hence:

$$f = \beta_1 + \beta_2 - \xi_{12} \quad (3)$$

$$f = \beta_1 = \beta_2 = \xi = 3 \quad (4)$$

Thus, the system can be expressed with $n = p - 3$ parameters in a dimensionless form which corresponds to the number of inputs required for the parameter scaling process.

*2.3. Identification of Parameters Influencing the Scaled Quantity*

In order to select the set of $n$ dimensionless numbers, the next step involves identifying which parameters have influence over the scaled parameter ($F_c$) based on the underlying physics (in this case dynamics) of the system. Figs. 3 and 4 demonstrate the free body diagrams of the bucket and the overall mechanism. In these figures, the component numbering from Table 1 is used. $F_{ab}$ represents the force at the joint connecting the components numbers $a$ and $b$, whereas $F_{ab_x}$ and $F_{ab_y}$ represent $x$ and $y$ components of $F_{ab}$ respectively.

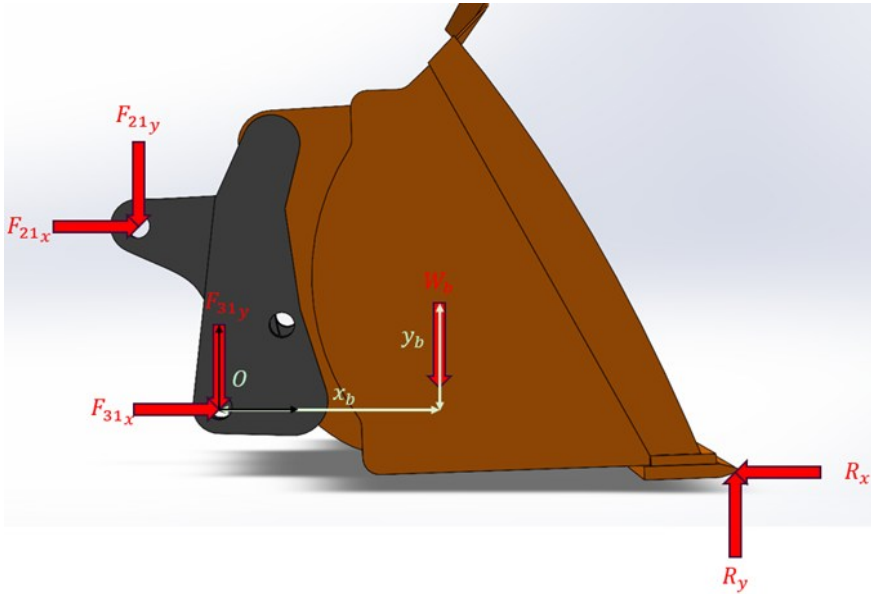

**Fig. 3.** Free body diagram of the bucket.

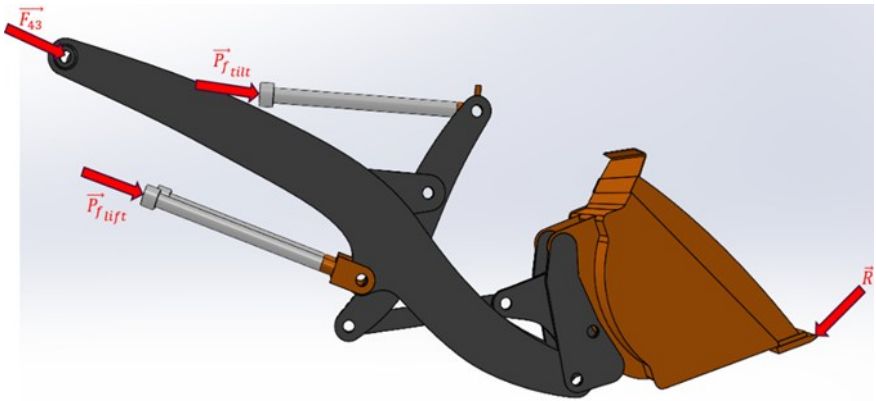

**Fig. 4.** Free body diagram of the overall mechanism.

Let $R$, $I_O$, $m_1$, $\alpha_1$, and $a_1$ denote, respectively, the excavation reaction force acting on the bucket, the mass moment of inertia about point $O$, the mass of the bucket, its angular acceleration, and the linear acceleration at its center of mass. The following equations of motion are obtained from the bucket's free body diagram:

$$\sum \vec{F} = m_1 \vec{a_1} \quad (5)$$

$$\sum \vec{F} = \vec{F_{21}} + \vec{F_{31}} \quad (6)$$

$$\sum \vec{M_O} = I_O \vec{\alpha_1} \quad (7)$$

$$\sum \vec{M_O} = \vec{r_{21}} \times \vec{F_{21}} + \vec{r_W} \times \vec{W} + \vec{r_R} \times \vec{R} \quad (8)$$

Let $P_{f_{tilt}}$ and $P_{f_{lift}}$ denote the forces generated by the pressures on the hydraulic cylinders responsible for tilting and lifting the buckets. Then, from the free body diagram of the overall mechanism:

$$\sum \vec{F} = \vec{F_{43}} + \vec{R} + \vec{P_{f_{tilt}}} + \vec{P_{f_{lift}}} \quad (9)$$

As illustrated in Fig. 5, while generating the training data of the scaling model, the simulation environment of the digital twin was modified so that the bucket and the load that it carries were treated as a combined dynamic body. In this configuration, both the mass and the center of mass location varied depending on the amount of load material loaded from the pile.

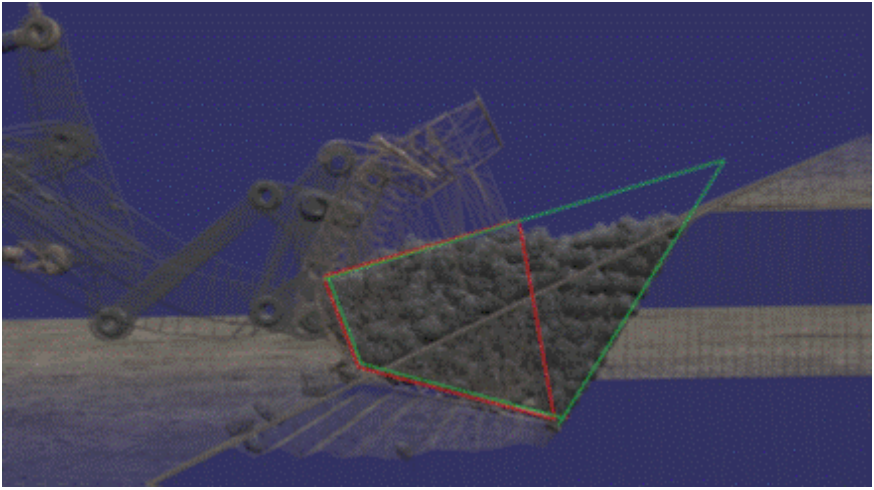

**Fig. 5.** The green area encompassing all terrain bodies generated in a given time, and the red area encompassing the material inside the bucket.

Based on the dynamics of the system, the following set of parameters in Eq. (10) were chosen to be included in the model that is designed to scale the target parameter, $F_{21}$ (magnitude of the force at the central joint load pin), which is being scaled under the influence of distorted scaling factors.

$$F_{21} = F_{21}\left(F_{31}, P_{f_{tilt}}, P_{f_{lift}}, x_b, y_b, m_b, a_1, \alpha_1\right) \quad (10)$$

where $x_b, y_b$, and $m_b$ represent the center-of-mass coordinates and the mass of the dynamic, combined body (bucket and load), respectively, on the rotating coordinate system attached to point $O$ (as shown in Fig. 3).

*2.4. Selection of the Input and Output Parameters*

Using Eq. (1), it can be observed that Eq. (10), which includes 9 parameters, can be reformulated in a dimensionless form involving only 6 parameters, as follows:

$$n = p - f = 9 - 3 = 6 \quad (11)$$

The dimensionless form of Eq. (10) is expressed as shown in Eq. (12).

$$\pi_1 = F(\pi_2, \pi_3, \pi_4, \pi_5, \pi_6) \quad (12)$$

where $\pi_n$ denotes dimensionless terms derived from parameters in the original form of Eq. (10). The term $\pi_1$ is the output of the model and it is deliberately chosen as to include the target parameter, $F_{21}$, so that its scaling can be facilitated.

$$\pi_1 = \frac{F_{21}}{F_{31}} \quad (13)$$

A Buckingham Pi theorem-based computer program was developed to generate a large number of dimensionless numbers based on the parameters that would be included in the model. These dimensionless numbers then went through another computer program which was developed to incorporate a correlation-based filtering mechanism so that a set of dimensionless terms which exhibit strong correlation with the parameter being scaled can be identified. The overall correlation strength of each set was evaluated based on the Pearson correlation coefficient [39,40] between the target parameter and each individual dimensionless term, followed by calculating the root-mean-square (RMS) of these individual correlation values.

*2.5. Inclusion of Distortion Effects on Scaling*

In an ideal scenario, where scaling can be achieved by a uniform scaling factor, the scaling laws can be defined in the following format [17]:

$$\pi_n^{\,m} = \pi_n^{\,p} \quad (14)$$

where the dimensionless numbers of one model are denoted with $\pi_n^{\,m}$ and those of the prototype are represented by $\pi_n^{\,p}$. When distorted scaling factors are involved, the scaling laws will instead adhere to the relationship described in Eq. (15) below:

$$\pi_n^{\,m} = \delta_n \pi_n^{\,p} \quad (15)$$

For the dimensionless term with the target parameter:

$$\pi_1^{\,m} = \delta_1 \pi_1^{\,p} \quad (16)$$

In Eq. (16), $\delta_1$ is referred to as the prediction factor. This factor is a function of the distortions and the remaining dimensionless numbers [17]. Here we develop a compact formulation to model $\delta_1$ in a manner that includes the effects of both the remaining dimensionless numbers and the scaling distortions. This relationship is expressed as shown in Eq. (17):

$$\delta_1 = \delta_1(d_2, d_3, d_4 \ldots d_n) \quad (17)$$

where each distortion term $d_n$ is defined by the ratio of the dimensionless number of the model to that of the prototype:

$$d_n = \frac{\pi_n^{\,m}}{\pi_n^{\,p}} \quad (18)$$

Since this system can be expressed using 6 dimensionless parameters, the prediction factor incorporating the target parameter being scaled can be modeled as follows:

$$\delta_1 = \delta_1(d_2, d_3, d_4, d_5, d_6) \quad (19)$$

Since $\pi_1^{\,m}$ is a known quantity in Eq. (16), which is obtained from the digital twin of the instrumented commercial wheel loader, $\pi_1^{\,p}$ can be evaluated using the prediction factor $\delta_1$ estimated by the scaling model. Once $\pi_1^{\,p}$ is determined, the scaled parameter for the prototype (in this case $F_{21}^{\,p}$), which is costly and challenging to measure directly using an additional load pin, can be estimated by substituting the known, physical parameters as follows:

$$\delta_1 = \frac{\pi_1^{\,m}}{\pi_1^{\,p}} \quad (20)$$

$$\pi_1^{\,p} = \frac{F_{21}^{\,p}}{F_{31}^{\,p}} \quad (21)$$

$$F_{21}^{\,p} = F_{31}^{\,p} \frac{\pi_1^{\,m}}{\delta_1} \quad (22)$$

Once the target parameter is accurately scaled by the model, the instrumented/calibrated digital twin of a unit in a product line can be leveraged to establish the digital twin of a different unit in the same product line with vastly distorted scaling factors without the need for additional instrumentation and calibration.

*2.6. Training Process of the Scaling Model*

Three wheel loaders were considered within the scope of this study. The calibrated digital twin of a medium-sized 3300 kg EVERUN ER12 wheel loader as well as the simulations of a roughly 26000 kg Komatsu WA475 wheel loader were utilized to generate the training data of the model and to verify it. The third wheel loader was chosen as a miniature 11 kg Kabolite wheel loader which was employed to validate the accuracy of the trained models under extreme scale differences with large degrees of scaling distortions. In each simulation, wheel loaders carry out a similar sequence of standardized actions where they move in a straight line, lower the bucket to collect material, and then pull back after completing the loading action. The evaluated parameters are recorded from the start to the end of the operation and incorporated into the training data. Three distinct models were trained to scale the target parameter. The first one employed feed-forward neural networks (FFNN) [41] utilizing the parameter set from Eq. (10). The second one was a Gated Recurrent Unit (GRU)- based [42] recurrent neural network (RNN) [43] with the same set of parameters. The last one also utilized RNNs but used an alternative parameter set, detailed below.

$$F_{21} = F_{21}\left(F_{31}, P_{f_{tilt}}, P_{f_{lift}}, x_b, y_b, m_b, V_1, w_1\right) \quad (23)$$

Unlike the FFNN-based model, the temporal sequence of the training data plays a critical role in capturing dynamic behavior for the models using RNNs. To preserve these temporal dependencies inherent in the simulation, the training data was organized into ordered sequences that reflect how the events progress within each simulation run. Each sequence was constructed such that the order in which data points were fed into the model corresponded to how the timeline progress in the simulations. The dimensionless term associated with the parameter that is intended to be scaled is set as the model's output ($\frac{\pi_1^m}{\pi_1^p}$, or $\frac{F_{21}^m F_{31}^p}{F_{31}^m F_{21}^p}$), while the rest of the dimensionless terms serve as inputs. The model whose results are scaled is designated randomly as one row from the training dataset corresponding to the digital twin of the ER12 medium-size wheel loader. Each modeling approach was evaluated based on its accuracy in scaling the target parameter. Furthermore, hyperparameter optimization was utilized to fine-tune key architectural and training parameters, including, the number of hidden layers [44], number of neurons per hidden layer [45], dropout rate [46], and the learning rate [47].

## 3. Results of the Parameter Scaling

Figs. 6 and 7 illustrate the performance of the models trained on the same dataset, evaluated using the coefficient of determination ($R^2$) [48]. In these figures, the values that models predict for $\delta_1$ are plotted against the real values of $\delta_1$. The y = x lines represent perfect alignment, serving as a reference for evaluating the model's performance. Fig. 6 shows the results of the FFNN-based model employing the set of parameters from Eq. (10). This model exhibited the best performance with an $R^2$ value of 0.99. Fig. 7 illustrates the performance of the RNN-based models where the one incorporating the set from Eq. (23) shown on the right and the one using the set from Eq. (10) shown on the left. These models yielded coefficient of determination values of 0.85 and 0.81 respectively, which indicates reduced predictive performance compared to the FFNN-based model.

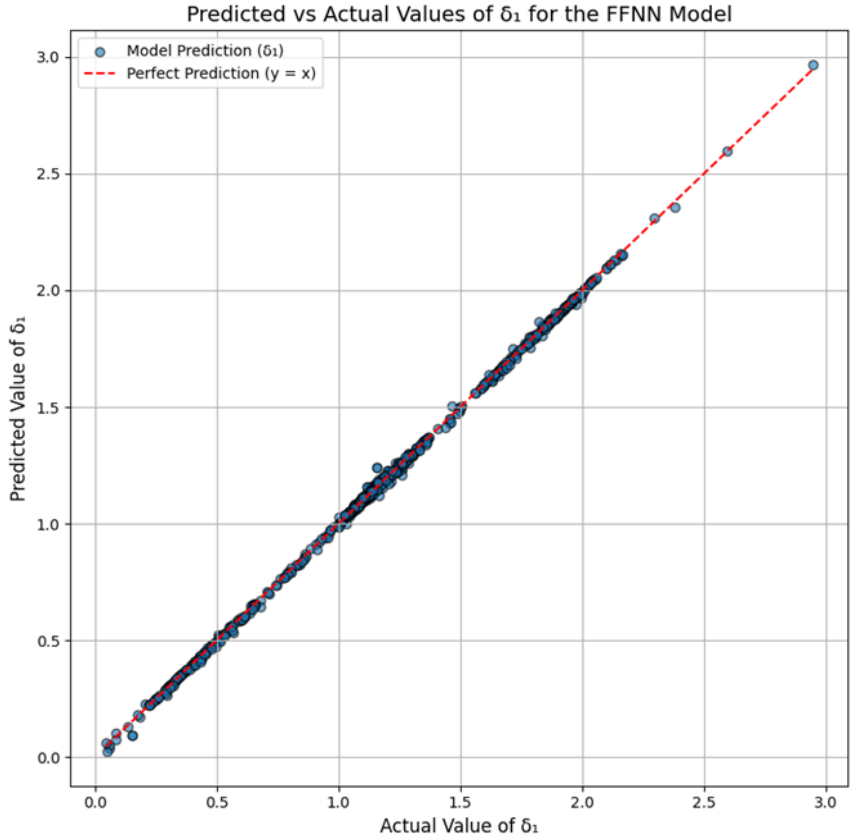

**Fig. 6.** Predicted values of the scaling law plotted against the actual values with the trendline representing perfect accuracy in scaling.

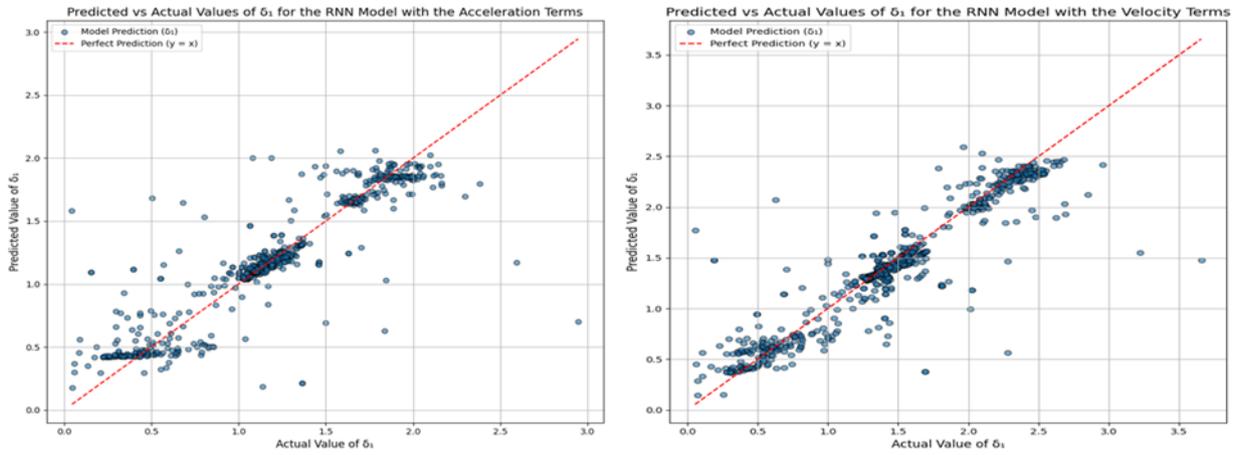

**Fig. 7.** Predicted values of the scaling law plotted against the actual values with the trendline representing perfect accuracy in scaling.

Figs. 8, 9, and 10 present the grid search process [49,50] conducted to optimize the hyperparameters of the FFNN-based model. Fig. 8 presents a two-dimensional heat map which depicts model performance as a function of the number of units per layer and dropout rate. Color in this figure indicates the mean $R^2$ value. As can be seen from the figure, configurations with narrow layers and high dropout rates exhibit significantly degraded performance whereas lower dropout rates combined with moderate to large numbers of units per layer achieved the best performance. Fig. 9 demonstrates a scatter plot of the hyperparameter landscape where the number of units per layer, dropout rate, and the resulting coefficient of determination represent the three axes. Color represents learning rate, and marker shape denotes the number of hidden layers. Fig. 10 shows how $R^2$ varies based on the number of units per hidden layer, averaged across all other hyperparameter combinations. The curve shows a sharp rise in performance at the beginning, going from 1 to 16 units. Beyond this point, the performance plateaus which indicates diminishing returns.

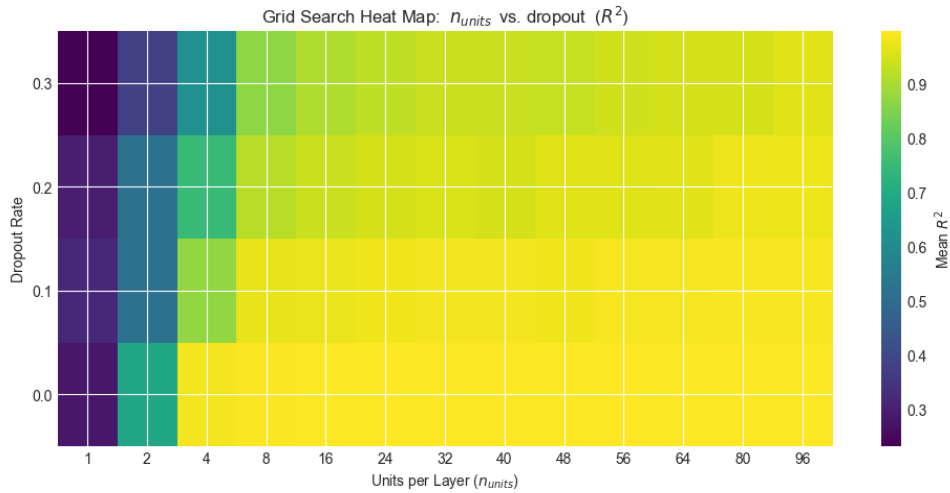

**Fig. 8.** Visualization of the grid search method, illustrating the impact of hidden layer width and dropout rate on model performance, measured by the mean $R^2$.

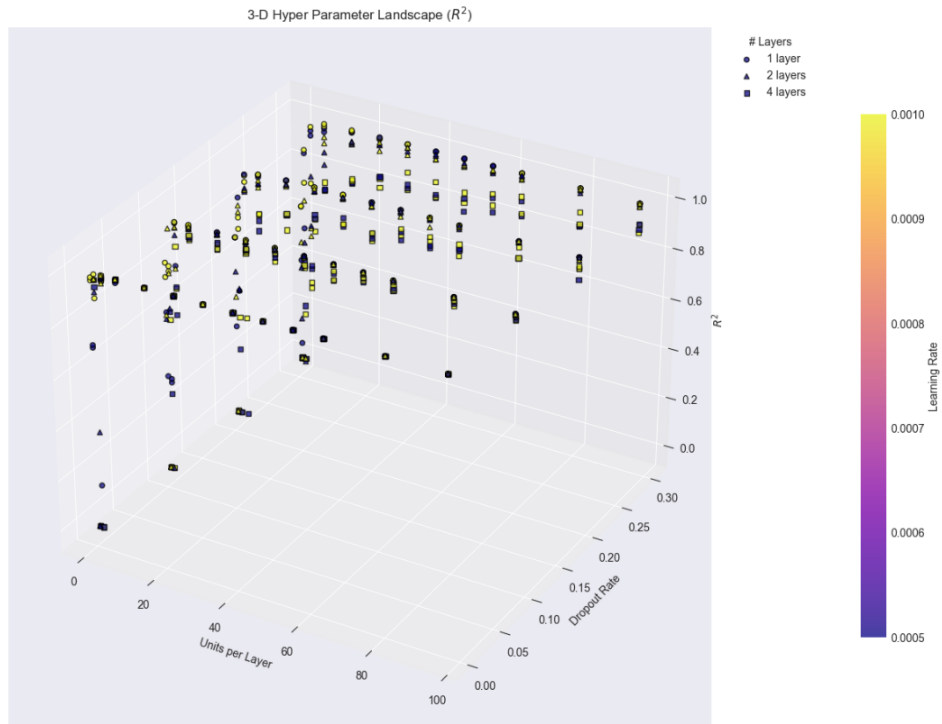

**Fig. 9.** Visualization of the grid search method, illustrating the hyperparameter landscape with the effect of hidden layer width, dropout rate, learning rate (color scale), and number of layers (marker shape) on model performance.

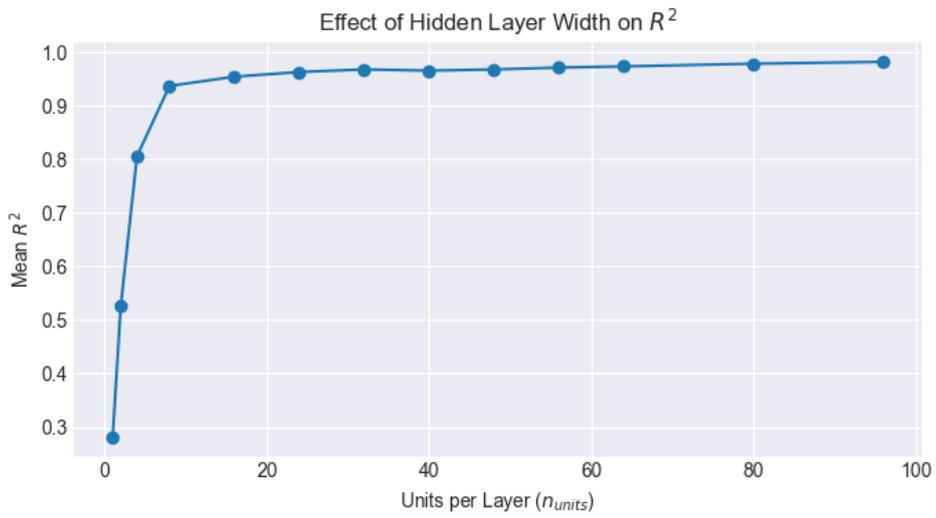

**Fig. 10.** Visualization of the grid search method, illustrating diminishing returns on accuracy beyond a certain network width.

## 4. Validation of the Scaling Model Accuracy on a Miniature Wheel Loader

The trained scaling framework was validated on a Kabolite miniature wheel loader displayed on Fig. 11. Table 2 presents a comparison of the estimated masses of the end loader mechanism components of the EVERUN ER12 wheel loader and the miniature wheel loader. The estimations for the full-scale wheel loader were obtained from its detailed CAD model which was constructed based on detailed measurements taken from the vehicle. The masses for the smaller wheel loader were estimated based on how the dimensions of the components scale.

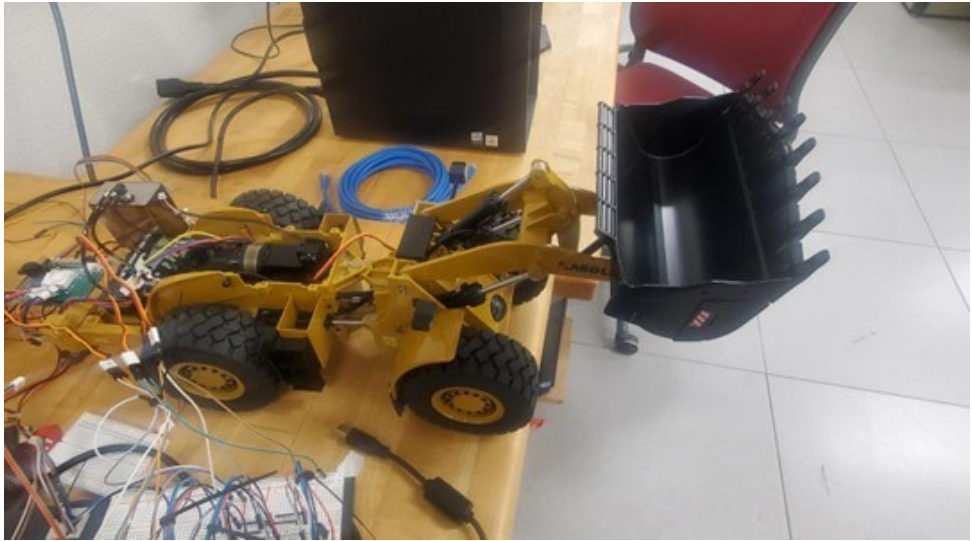

**Fig. 11.** Miniature size wheel loader on a desk.

**Table 2**
Masses of the end loader mechanism components for the full-size EVERUN ER12 wheel loader and the miniature wheel loader.

| Part - Full-size wheel loader | Mass (kg) | Part - Miniature-size wheel loader | Mass (kg) |
| --- | --- | --- | --- |
| Bucket | 205.8 | Bucket | 0.23 |
| Bucket Base | 84.8 | Bucket Base | 0.09 |
| Link B | 8.99 | Link B | 0.01 |
| Link A | 33.9 | Link A | 0.04 |
| Main Arm | 294.6 | Main Arm | 0.32 |
| Hydraulic Rod A1 | 20.3 | Hydraulic Rod A1 | 0.02 |
| Hydraulic Rod A2 | 13.1 | Hydraulic Rod A2 | 0.01 |
| Hydraulic Rod B1 | 19.6 | Hydraulic Rod B1 | 0.02 |
| Hydraulic Rod B2 | 14.7 | Hydraulic Rod B2 | 0.02 |

Unlike the EVERUN ER12 wheel loader, the miniature wheel loader does not feature an extensive set of sensors. Consequently, data for the miniature wheel loader was generated by assuming a static equilibrium condition under known weights, enabling an analytical solution for the required parameters. The model is exclusively trained using data from the Everun and Komatsu wheel loaders, while data from the miniature wheel loader is utilized for validation. The scaling models were evaluated under conditions where the load carried by the miniature wheel loader ranged from 0.5 to 4 kilograms. The upper limit was set to 4 kg by proportionally scaling down the maximum load capacity of the full-sized wheel loader relative to the overall vehicle weight of the miniature model. Fig. 12 illustrates the evaluation of the scaling models on the miniature wheel loader where the x-axis represents the load that the miniature wheel loader carries, while the y-axis represents the percentage error in the model's estimations. Each curve corresponds to a different scaling model, and the dashed lines denote the average percentage error for each model across the entire load range. To establish a baseline for comparison, the direct application of the Buckingham Pi theorem was employed. As anticipated, since this approach does not account for scaling factor distortions, it resulted in a significantly high average error of 42.04%. The best-performing scaling model was the FFNN-based model which resulted in an average error of 4.29%. This represents an approximate 90% reduction in error compared to the direct application of the Buckingham Pi theorem, which does not account for scaling distortions. This result demonstrates that it is feasible to use the scaling framework to reduce the number of sensors needed for digital twin calibration. In this context, leveraging scalability greatly reduces the cost of developing complex, physics-based digital twins.

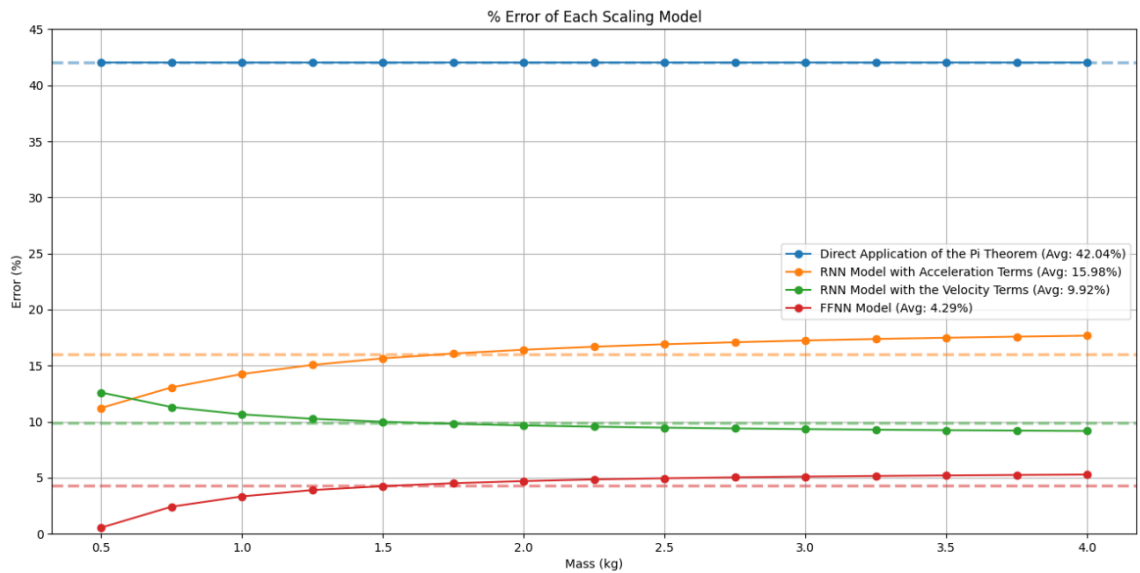

**Fig. 12.** Errors in the scaling models' estimations for the miniature wheel loader carrying known weights.

## 5. Conclusions

This paper presented the development of a scaling framework for digital twin models and validated its accuracy across wheel loaders of vastly different scales. The scaling framework integrated a novel methodology and modular computational programs that utilize dimensional analysis and machine learning to facilitate the scaling process. The developed framework's ability to account for the effects of distorted scaling factors on parameter scaling enables the use of small-scale units in similar production lines to be utilized in developing digital twins of larger-scale units without the need for calibration. This would not be possible with the conventional scaling methods based solely on the Buckingham Pi theorem. The integration of machine learning was pivotal, particularly when it comes to capturing how distortions impact parameter scaling.

**CRediT authorship contribution statement**

**Deniz Karanfil:** Writing – original draft, Conceptualization, Visualization, Validation, Methodology, Investigation, Formal analysis, Resources, Data curation, Software.

**Bahram Ravani:** Writing – review & editing, Project administration, Investigation, Supervision, Funding acquisition, Conceptualization, Validation, Methodology.

**Acknowledgments**

Funding: This work was supported in part by funding from Komatsu of Japan. Their support for this work is greatly appreciated.